\documentclass[reprint,superscriptaddress,amsmath,amssymb,aps,pra]{revtex4-1}
\usepackage{graphicx}
\usepackage{dcolumn}
\usepackage{bm}
\usepackage{tabularx}
\usepackage{amsmath}
\usepackage{xcolor}
\usepackage{hyperref}

\hypersetup{
    colorlinks=true,
    linkcolor=blue,
    filecolor=blue,
    urlcolor=blue,
    citecolor=blue
}

\emergencystretch=\maxdimen
\begin{document}

\title{Evidence for multiband gapless superconductivity in the topological superconductor candidate 4Hb-TaS\texorpdfstring{$_{2}$}{}}

\author{Hanru Wang}
  \thanks{These authors contributed equally to this work.}
  \affiliation{State Key Laboratory of Surface Physics, Department of Physics, Fudan University, Shanghai 200438, China}
  \affiliation{Shanghai Research Center for Quantum Sciences, Shanghai 201315, China}

\author{Yihan Jiao}
  \thanks{These authors contributed equally to this work.}
  \affiliation{State Key Laboratory of Surface Physics, Department of Physics, Fudan University, Shanghai 200438, China}

\author{Fanyu Meng}
  \thanks{These authors contributed equally to this work.}
  \affiliation{School of Physics and Beijing Key Laboratory of Opto-electronic Functional Materials $\&$ Micro-nano Devices,\\
    Renmin University of China, Beijing 100872, China}
  \affiliation{Key Laboratory of Quantum State Construction and Manipulation (Ministry of Educatoin), \\
    Renmin University of China, Beijing 100872, China}

\author{Xu Zhang}
  \affiliation{State Key Laboratory of Surface Physics, Department of Physics, Fudan University, Shanghai 200438, China}

\author{Dongzhe Dai}
  \affiliation{State Key Laboratory of Surface Physics, Department of Physics, Fudan University, Shanghai 200438, China}

\author{Chengpeng Tu}
  \affiliation{State Key Laboratory of Surface Physics, Department of Physics, Fudan University, Shanghai 200438, China}

\author{Chengcheng Zhao}
  \affiliation{State Key Laboratory of Surface Physics, Department of Physics, Fudan University, Shanghai 200438, China}

\author{Lu Xin}
  \affiliation{State Key Laboratory of Surface Physics, Department of Physics, Fudan University, Shanghai 200438, China}

\author{Sicheng Huang}
  \affiliation{State Key Laboratory of Surface Physics, Department of Physics, Fudan University, Shanghai 200438, China}

\author{Hechang Lei}
  \email{Contact author: hlei$@$ruc.edu.cn}
  \affiliation{School of Physics and Beijing Key Laboratory of Opto-electronic Functional Materials $\&$ Micro-nano Devices,\\
    Renmin University of China, Beijing 100872, China}
  \affiliation{Key Laboratory of Quantum State Construction and Manipulation (Ministry of Educatoin), \\
    Renmin University of China, Beijing 100872, China}

\author{Shiyan Li}
  \email{Contact author: shiyan$\_$li$@$fudan.edu.cn}
  \affiliation{State Key Laboratory of Surface Physics, Department of Physics, Fudan University, Shanghai 200438, China}
  \affiliation{Shanghai Research Center for Quantum Sciences, Shanghai 201315, China}
  \affiliation{Shanghai Branch, Hefei National Laboratory, Shanghai 201315, China}
  \affiliation{Collaborative Innovation Center of Advanced Microstructures, Nanjing 210093, China}

\date{\today}

\begin{abstract}

    We present the ultralow-temperature thermal conductivity measurements on single crystals of transition-metal dichalcogenide material 4Hb-TaS$_{2}$, which has recently been proposed as a topological superconductor candidate.
    In zero field, a small residual linear term $\kappa_{0}/T$ is observed, indicating the existence of a residual density of states in the superconducting state. The slow field dependence of $\kappa_{0}/T$ at low fields rules out the presence of nodes in the superconducting gap, and the S-shaped field dependence across the full field range suggests multiple superconducting gaps in 4Hb-TaS$_{2}$.
    Our results provide evidence for multiband gapless superconductivity in 4Hb-TaS$_{2}$, and the residual density of states come from certain gapless Fermi surfaces.

\end{abstract}

\maketitle

Topological superconductor (TSC) has emerged as a pivotal topic in condensed matter physics due to their ability to host Majorana zero modes (MZMs) \cite{Qi2011,Sato_2017,Xu2019}. MZMs obey non-Abelian statistics and could be utilized to realize the fault-tolerant quantum computing \cite{RevModPhys.80.1083}. The chiral superconductor is considered  one of the most promising platforms for realizing intrinsic topological superconductivity \cite{Kallin_2016}. For 2D chiral TSCs, the simplest model is the chiral $p_{x}+ip_{y}$ pairing state of spinless fermions, which predicts chiral Majorana edge modes and a single Majorana zero mode in a superconducting vortex \cite{Green2000,Ivanov2001}. Up to now, chiral superconductors have rarely been observed experimentally, with the main candidates being Sr$_{2}$RuO$_{4}$ \cite{Kallin_2012,Mackenzie2003,Maeno2012}, UPt$_{3}$ \cite{Louis2002,Schemm2014} and UTe$_{2}$ \cite{Ran2019,Ishihara2023,Hayes2021}. However, the latest nuclear magnetic resonance (NMR) experiments have ruled out spin-triplet pairing in Sr$_{2}$RuO$_{4}$ \cite{Pustogow2019}. Therefore, it is crucial to identify more intrinsic topological superconductors and thoroughly investigate their superconducting pairing nature.

The transition-metal dichalcogenide (TMD) material 4Hb-TaS$_{2}$ has recently attracted significant attention due to its natural van der Waals heterostructure (vdWH), which is crystallized in a layered hexagonal lattice and consists of alternating layers of 1T-TaS$_{2}$ and 1H-TaS$_{2}$, as shown in Fig. 1(a).
Monolayer 1T-TaS$_{2}$ is a strongly correlated Mott insulator and was proposed to be a quantum spin liquid (QSL) candidate \cite{Patrick2018}.
Meanwhile, monolayer 1H-TaS$_{2}$ exhibits Ising superconductivity with a superconducting transition temperature $T_{\rm c} = 3.4 \ $K \cite{monolayer1H2018,Yang2018}.
Muon spin relaxation ($\mu$SR) measurements reported the signature of time-reversal symmetry breaking below $T_{\rm c}$, suggesting that 4Hb-TaS$_{2}$ is a chiral superconductor candidate \cite{Ribak2020}. Furthermore, recent scanning tunneling microscopy (STM) experiments have observed topological edge states in 4Hb-TaS$_{2}$, indicating it could be a natural topological superconductor \cite{Beidenkopf2021}.
Two-component nematic superconductivity \cite{Silber2024, littlepark2022} and the finite momentum pairing superconducting states \cite{Yang2024} have also been proposed.
Additionally, the magnetic memory effect observed in 4Hb-TaS$_{2}$ suggests an unusual magnetic phase above $T_{\rm c}$ \cite{Kalisky2022}, which may favor an interlayer spin-triplet pairing state \cite{Altman2024}.
All these results establish 4Hb-TaS$_{2}$ as an ideal system for exploring topological superconductivity, unconventional magnetism and their interplay within the monolayer limit.

Clarifying the superconducting gap structure will provide valuable insights into the superconducting pairing mechanism.
Both specific heat and transverse-field $\mu$SR measurements showed a fully gapped superconductivity in 4Hb-TaS$_{2}$ \cite{Ribak2020}.
However, scanning tunneling spectroscopy (STS) measurements on monolayer 1H-TaS$_{2}$ \cite{Liljeroth2023} and on 1H layer of 6R-TaS$_{2}$ (another stacking of 1T-TaS$_{2}$ and 1H-TaS$_{2}$) \cite{6R2024} have revealed a V-shaped nodal-like superconducting gap.
More interestingly, a residual linear term in the electronic specific heat was found in the superconducting state of 4Hb-TaS$_{2}$, with the origin unknown \cite{Ribak2020}.
A recent theory attributed it to the gapless superconductivity displaying on part of the Fermi surfaces \cite{Dentelski2021}.
Finite in-gap density of states (DOS) were also observed in STS experiments \cite{Beidenkopf2021,6R2024}, which was attributed to topological nodal-point superconductivity associated with the inter-orbital pairing channel \cite{Beidenkopf2021}.
Therefore, the superconducting gap structure of 4Hb-TaS$_{2}$ remains unclear, highlighting the need for further experimental investigation.

In this Letter, we report the ultralow-temperature thermal conductivity measurements on 4Hb-TaS$_{2}$ single crystals.
In  zero  field, a small residual linear term $\kappa_{0}/T$ is observed, suggesting the presence of residual DOS in the superconducting state.
The field dependence of $\kappa_{0}/T$ exhibits a slow increase at low field, followed by an S-shaped curve arcoss the entire field range, indicating multiple nodeless superconducting gaps.
Based on these experimental results, we infer that 4Hb-TaS$_{2}$ likely hosts multiband gapless superconductivity, i.e., with some Fermi surfaces fully gapped and certain Fermi surfaces gapless.

\begin{figure}
\includegraphics[clip,width=8.5cm]{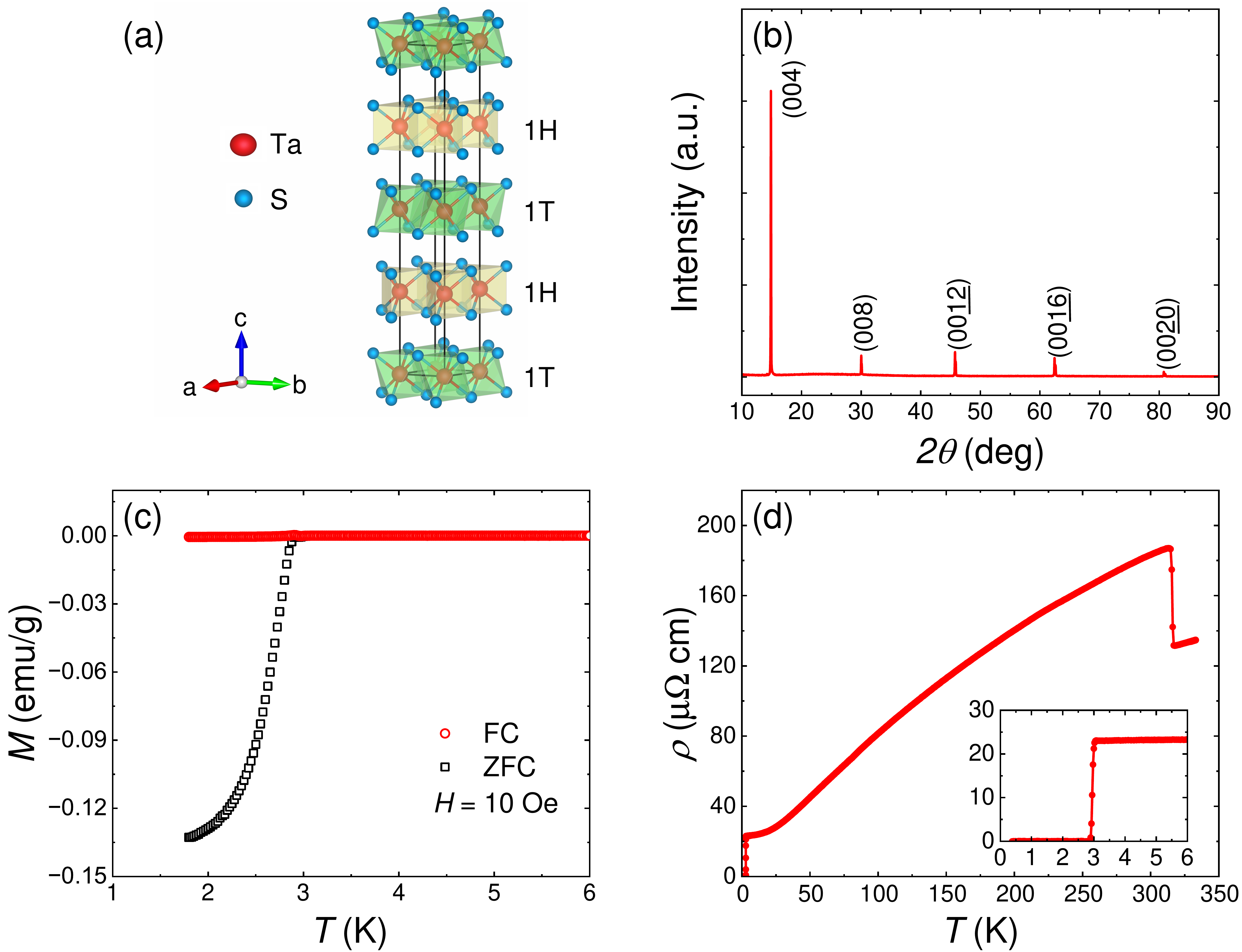}
\caption{(a) Crystal structure of 4Hb-TaS$_{2}$. (b) The X-ray diffraction pattern for the largest natural surface of 4Hb-TaS$_{2}$ single crystal. Only (00$l$) Bragg peaks were observed. (c) Low temperature dc magnetization of a 4Hb-TaS$_{2}$ single crystal at $H$ = 10 Oe, parallel to the $ab$ plane, with both zero-field-cooling (ZFC) and field-cooling (FC) processes. (d) In-plane resistivity of 4Hb-TaS$_{2}$ single crystal in zero field. The inset shows the superconducting transition at low temperature.}
\end{figure}

Single crystals of 4Hb-TaS$_{1.99}$Se$_{0.01}$ were grown by the chemical vapor transport method \cite{Gao2020}.
A tiny amount of Se doping enhances the stability of the crystal structure and significantly increases the superconducting volume fraction, which has been done in most previous studies \cite{Beidenkopf2021,Ribak2020,Kalisky2022,littlepark2022,Silber2024,Lei2024,Almoalem2024,Nayak2023}.
The X-ray diffraction (XRD) measurement was performed on an X-ray diffractometer (D8 Advance, Bruker). The largest natural surface of 4Hb-TaS$_{2}$ single crystals was determined to be the (00$l$) plane, as illustrated in Fig. 1(b).
The dc magnetization measurement was performed down to 1.8 K using a magnetic property measurement system (MPMS, Quantum Design).
The specific heat and in-plane resistivity were measured in a physical property measurement system (PPMS, Quantum Design) equipped with a ${ ^3}$He cryostat.
Samples for resistivity and thermal conductivity measurements were cut into rectangular shapes. The dimensions were 2.68 $\times$ 0.61 $\times$ 0.071 mm$^3$ for Sample A and 1.70 $\times$ 0.89 $\times$ 0.062 mm$^3$ for Sample B, respectively.
Four silver wires were attached to the sample with silver paint. The thermal conductivity was measured in a dilution refrigerator, using a standard four-wire steady-state method with two RuO$_{2}$ chip thermometers, calibrated $in$ $situ$ against a reference RuO$_{2}$ thermometer.
To ensure a homogeneous field distribution in the sample, all fields were applied at a temperature above $T_{\rm c}$ for transport measurements.

\begin{figure}
  \includegraphics[clip,width=7.0cm]{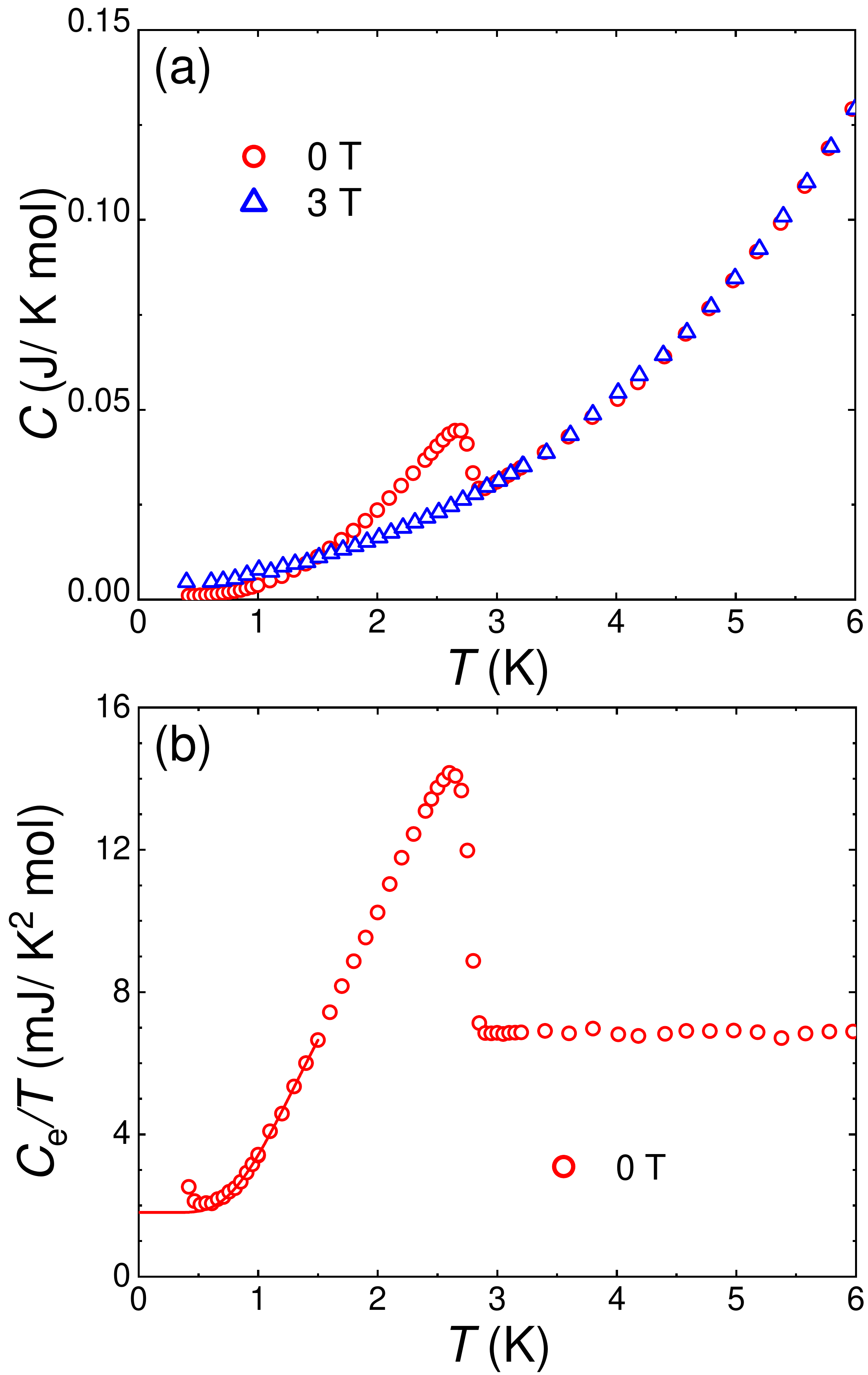}
  \caption{(a) Temperature dependence of the specific heat for 4Hb-TaS$_{2}$ single crystal in zero field and 3 T. (b) Zero-field electronic specific heat $C_{\rm e}$ obtained by subtracting the phonon specific heat, plotted as $C_{\rm e}/T$ vs $T$. The red line is a fit of the data between 0.8 and 1.5 K to $C_{e}/T = \gamma_{r}+\frac{A}{T}exp(-\Delta/kT)$. A linear term $\gamma_{r}$ = 1.80 $\pm$ 0.07 mJ K$^{-2}$ mol$^{-1}$ is obtained, accounting for $26\%$ of the total electronic specific heat in the normal state.}
\end{figure}

Figure 1(c) plots the low-temperature dc magnetization of a 4Hb-TaS$_{2}$ single crystal at 10 Oe ($\mu_{0} H$ $||$ $ab$) with both zero-field-cooling (ZFC) and field-cooling (FC) processes, showing a clear diamagnetic superconducting transition at $T_{\rm c}\approx$ 2.9 K.
Figure 1(d) presents the in-plane resistivity of 4Hb-TaS$_{2}$ Sample A from 0.35 to 330 K. The sharp anomaly at 315 K is due to the $\sqrt{13} \times \sqrt{13}$ commensurate charge density wave (CCDW) transition in the 1T layer \cite{Kim1995}.
As shown in the inset of Fig. 1(d), the $T_{\rm c}$ defined by $\rho=0$ is 2.9 K, in agreement with the magnetization measurement.
The normal-state resistivity from 3 to 50 K can be well fitted by the Fermi liquid behavior $\rho\left(T\right)=\rho_{0}+AT^2$, with the residual resistivity $\rho_{0} = 22.8 \ \mu \Omega$ cm. The residual resistivity ratio $RRR =\rho(300 \ \rm K)/\rho_{0}$ is 8.1.

The specific heat of 4Hb-TaS$_{2}$ single crystal in zero field and 3 T is displayed in Fig. 2(a).
In zero field, a clear jump near $T_{\rm c}$ is observed. Above $T_{\rm c}$, the zero-field data from 3 to 6 K can be well fitted by $C = \gamma_{n} T +\beta T^3+\delta T^5$, which gives the electronic specific-heat coefficient $\gamma_{n} =6.86  \  {\rm mJ}\  {\rm K}^{-2}\ {\rm mol}^{-1}$, the phononic coefficient $\beta=0.378 \ {\rm mJ} \ {\rm K}^{-4}\ {\rm mol}^{-1}$ and $\delta=0.0009 \ {\rm mJ} \ {\rm K}^{-6}\ {\rm mol}^{-1}$.
After subtracting the phonon contribution, given by $\beta T^3+\delta T^5$, the electronic specific heat $C_{\rm e}$ is depicted in Fig. 2(b) as $C_{\rm e}/T$ versus $T$.
In zero field, $C_{\rm e}/T$ exhibits an upturn below 0.5 K. A similar upturn was also observed in bulk 1T-TaS$_{2}$ and possibly originates from the nuclear Schottky anomaly \cite{Ribak2017}.
$C_{\rm e}/T$ between 0.8 and 1.5 K can be well fitted by the  formula $C_{\rm e}/T = \gamma_{r} + \frac{A}{T} exp(-\Delta/kT)$, where the data is essentially unaffected by the low-temperature upturn.
The fitting gives an energy gap $\Delta$ = 0.39 meV and a finite residual linear term $\gamma_{r} = 1.80 \  {\rm mJ} \  {\rm K}^{-2} \  {\rm mol}^{-1}$.
The magnitude of the superconducting gap $\Delta$ is similar to that observed in $\mu$SR and STM measurements \cite{Beidenkopf2021,Ribak2020}, demonstrating the validity of the fitting.
The observation of finite residual linear term $\gamma_{r}$ is consistent with previous study \cite{Ribak2020}, which will be discussed late.

\begin{figure}
  \includegraphics[clip,width=7.0cm]{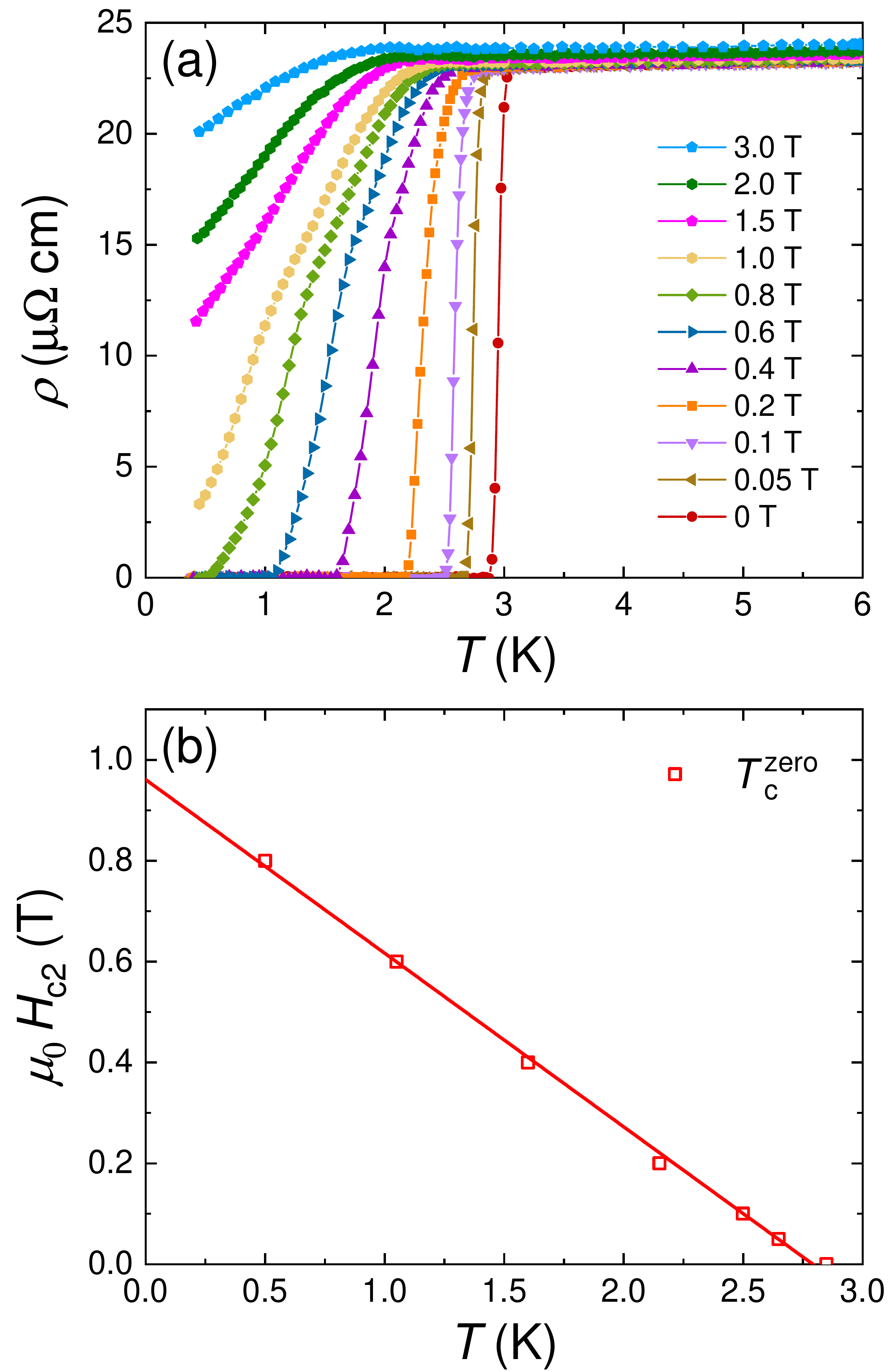}
  \caption{(a) Low-temperature in-plane resistivity of 4Hb-TaS$_{2}$ Sample A in magnetic fields up to 3 T.
  (b) Temperature dependence of the upper critical field $\mu_{0}H{\rm_{c2}}$, extracted from the $T^{\rm{zero}}_{\rm c}$ values in (a). The red line is a linear fit of $\mu_{0}H{\rm_{c2}}$, which gives $\mu_{0}H{\rm_{c2}}(0)\approx $ 0.96 T.}
  \end{figure}

To determine the upper critical field $\mu_{0} H_{\rm c2}(0)$ of 4Hb-TaS$_{2}$, we measured the low-temperature in-plane resistivity of Sample A in various magnetic fields ($\mu_{0} H \parallel c$) up to 3 T, as shown in Fig. 3(a). The temperature dependence of $H_{\rm c2} (T)$, defined by $\rho  = 0$ in Fig. 3(a), is plotted in Fig. 3(b). The red line is a linear fit to $\mu_{0} H_{\rm c2} (T)$, which can be described by the three-dimensional (3D) anisotropic Ginzburg-Landau (GL) theory \cite{Tinkham},

\begin{equation}
    \mu_{0} H_{\rm c2} (T)=\frac{\phi_{0}}{2\pi\xi_{\rm GL} (T)^2}=\frac{\phi_{0}}{2\pi \xi_{\rm GL}(0)^2}\left(1-\frac{T}{T_{\rm c}}\right),
\end{equation}
where $\phi_{0} =2.07\times 10^{-7}$  Oe cm$^2$ is the magnetic flux quantum, and $\xi_{\rm GL} (0)$ is the zero-temperature in-plane GL coherence length. The fit gives $\mu_{0} H_{\rm c2}(0) \approx 0.96\ \rm T$ and $\xi_{\rm GL}(0) \approx$ 18.5 nm.

Compared to specific heat experiments, ultralow-temperature thermal conductivity measurement eliminates the influence of nuclear Schottky effect, thereby providing more convincing evidence for low-lying excitations  and detailed information about the superconducting gap structure \cite{Shakeripour2009}.
In Fig. 4(a), we present the in-plane thermal conductivity of two 4Hb-TaS$_{2}$ single crystals in zero field.
The thermal conductivity at very low temperature can usually be fitted to $\kappa/T = a+bT^{\alpha-1}$, in which the two terms $aT$ and $bT^\alpha$ represent contributions from electrons and phonons, respectively \cite{Li2008,Sutherland2003}.
The residual linear term $\kappa_{0}/T \equiv a$ can be obtained by extrapolating $\kappa/T$ to $T=0$.
Because of the specular reflections of phonons at the sample surfaces, the power $\alpha$ in the second term is typically between 2 and 3 \cite{Li2008,Sutherland2003}.
In zero field, the fitting below 0.4 K gives $(\kappa_{0}/T)_{\rm A} = 24 \pm 4$ $\mu$W K$^{-2}$ cm$^{-1}$ for Sample A and $(\kappa_{0}/T)_{\rm B} = 24 \pm 3$ $\mu$W K$^{-2}$ cm$^{-1}$ for Sample B, with $\alpha_{\rm A}=2.49  \pm  0.04$  and $\alpha_{\rm B}=2.54   \pm  0.05$, respectively.
Comparing with our experimental uncertainty $\pm 5 \ \mu$W K$^{-2}$  cm$^{-1}$, this $\kappa_{0}/T$ value is not negligible.
For Sample A, according to the Wiedemann-Franz law, the expected value for the normal-state $\kappa_{\rm N0}/T$ is calculated as $L_{0}/\rho_{0} (0.96 \rm T)=1.065$ mW K$^{-2}$  cm$^{-1}$, where $L_{0} = 2.45 \times 10^{-8} \  \rm{W \ \Omega  \ K}^{-2}$ is the Lorenz ratio and $\rho_{0} (0.96 \rm T) = 23.0\ \mu \Omega$ cm.
The ratio $(\kappa_{0}/T)/(\kappa_{\rm N0}/T)$ for Sample A is about 2.3\%.
Similarly, for Sample B, the corresponding ratio is  2.7\%.
The finite residual linear term in thermal conductivity strongly supports the presence of residual DOS in the superconducting state of 4Hb-TaS$_{2}$.

A natural explanation for the residual DOS in superconducting states is the nodal quasiparticles from a nodal superconducting gap.
In conventional $s$-wave superconductors, since all electrons form Cooper pairs, there are no fermionic quasiparticles to conduct heat as $T \rightarrow 0$.
Therefore, there is no residual linear term $\kappa_{0}/T$, as seen in Nb \cite{Lowell1970} and NbSe$_{2}$ \cite{Boaknin2003}.
On the contrary, for nodal superconductors, a substantial $\kappa_{0}/T$ in zero field contributed by nodal quasiparticles has been found.
For instance, the $\kappa_{0}/T$ of the overdoped $d$-wave cuprate superconductor $\rm Tl_{2} Ba_{2} CuO_{6+\delta}$ (Tl-2201, $T_{\rm c} = 15 $ K) is 1.41 mW K$^{-2}$ cm$^{-1}$, $\sim $36\% $\kappa_{\rm N0}/T$ \cite{Proust2002}.

\begin{figure}
  \includegraphics[clip,width=7.2cm]{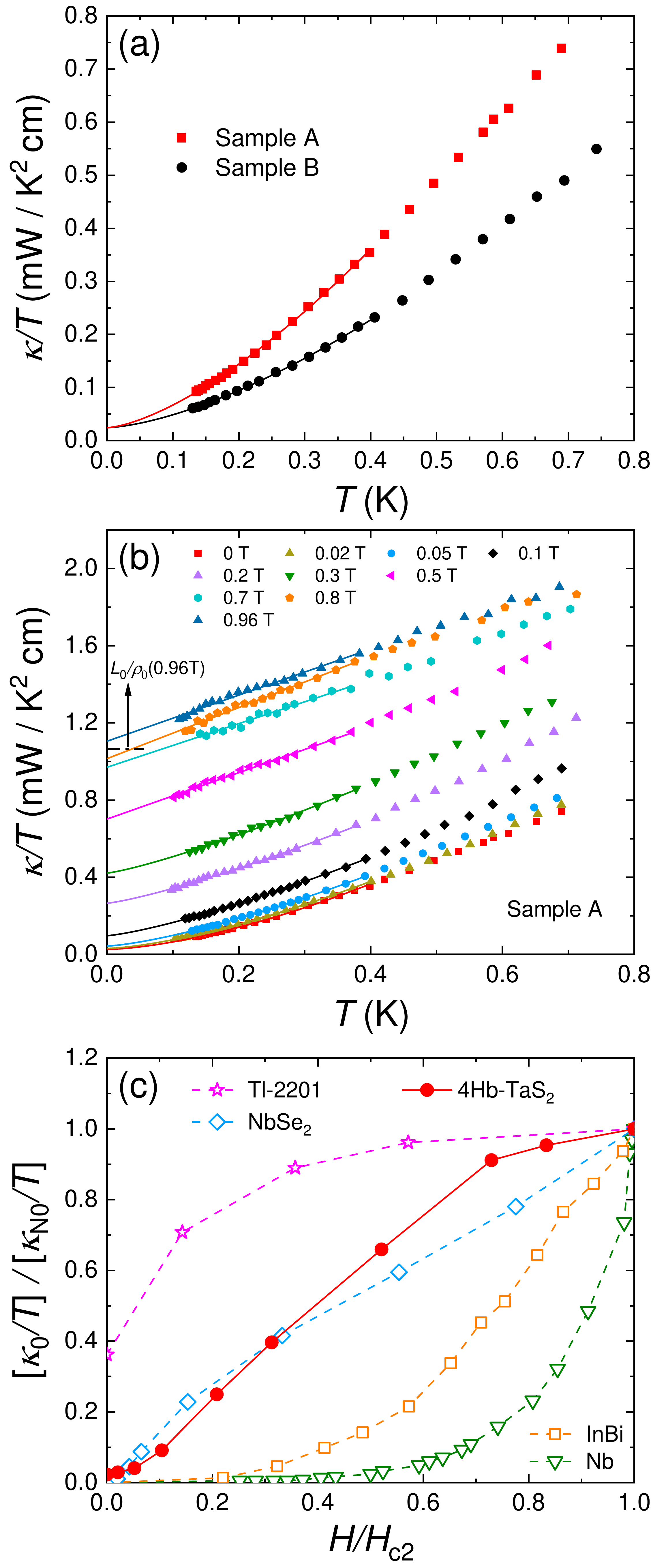}
  \caption{(a) Temperature dependence of the in-plane thermal conductivity for 4Hb-TaS$_{2}$ single crystals in zero field. The solid line represents a fit of the data below 0.4 K to $\kappa /T =a+bT^{\alpha-1}$, giving the residual linear term $(\kappa_{0} /T)_{\rm A} =24 \pm 4 \ \mu \rm W$  K$^{-2}$ cm$^{-1}$ for Sample A and $(\kappa_{0} /T)_{\rm B} =24 \pm 3 \ \mu \rm W$ K$^{-2}$ cm$^{-1}$ for Sample B, respectively.
  (b) Low-temperature thermal conductivity of Sample A in magnetic fields up to 0.96 T ($\mu_{0} H \parallel c$). The black dashed line is the normal-state Wiedemann-Franz law expectation $L_{0}/\rho_{0}(0.96 \rm \ T)$, with Lorenz number $L_{0} = 2.45 \times 10^{-8}$ W $\Omega$ K$^{-2}$ and $\rho_{0}\left(0.96 \rm \ T\right) =23.0 \  \mu\Omega$ cm.
  (c) Normalized $\kappa_{0}/T$ in (b) as a function of $H/H_{\rm c2}$. Comparative data from a clean $s$-wave superconductor Nb \cite{Lowell1970}, a dirty $s$-wave alloy InBi \cite{PhysRevB.14.1916}, a multiband $s$-wave NbSe$_{2}$ \cite{Boaknin2003}, and an overdoped $d$-wave cuprate Tl-2201 \cite{Proust2002} are also presented.}
\end{figure}

However, compared to Tl-2201, the residual $\kappa_{0}/T$ in 4Hb-TaS$_{2}$ is significantly smaller.
In fact, if the superconducting gap exhibits line nodes imposed by symmetry, the magnitude of $\kappa_{0}/T$ is expected to be a universal value, independent of the impurity scattering rate.
The universal conductivity has been validated in various nodal superconductors, such as  Tl-2201 \cite{Proust2002} and YBa$_{2}$Cu$_{3}$O$_{6.9}$ \cite{YBaCuO1997}.
For a quasi-2D $d$-wave superconductor \cite{universal}, the universal thermal conductivity satisfies

\begin{equation}
    \frac{\kappa_{0}}{T} \approx \frac{\hbar}{2\pi}\frac{\gamma_{\rm N}v_{\rm F}^2}{\Delta_{0}},
\end{equation}

where $\gamma_{\rm N}$ is the linear term in the normal-state electronic specific heat, $v_{\rm F}$ is the Fermi velocity, and $\Delta_{0}$ stands for the maximum of the superconducting gap.
For 4Hb-TaS$_{2}$, $\gamma_{\rm N} =6.86$  mJ K$^{-2}$  mol$^{-1}$ and $\Delta_{0}=0.39$ meV can be obtained from our data.
Meanwhile, using the relationship $\xi_{\rm GL}=\hbar v_{\rm F}/\pi \Delta_{0}$, we determine $v_{\rm F} $ to be $ 3.44 \times 10^4 $ m/s.
Assuming 4Hb-TaS$_{2}$ is a quasi-2D $d$-wave superconductor, we estimate $\kappa_{0}/T \approx 0.528 $ mW K$^{-2}$  cm$^{-1}$, which is about 50\% of $\kappa_{\rm N0}/T$.
This value significantly exceeds what we observed in our samples, thus making the scenario of line nodes unlikely.

We also notice the scenario of topological nodal-point superconductivity proposed in Ref. \cite{Beidenkopf2021}.
In case the gap vanishes at point nodes rather than line nodes, the residual $\kappa_{0}/T$ increases with the level of impurity scattering \cite{Shakeripour2009}.
At high scattering rate, $\kappa_{0}/T$ should become a substantial fraction of the normal-state $\kappa_{\rm N0}/T$ \cite{Graf1996}.
According to the Drude model, the normal-state impurity scattering rate $\Gamma_{0}$ can be roughly estimated from $\rho_{0}$ and the plasma frequency $\omega_{p} = c/ \lambda_{0} $, where $\lambda_{0}$ is the penetration depth, approximately  487 nm \cite{Ribak2020}.
We estimate $\hbar \Gamma_{0} / k_{\rm B} T_{\rm c} \approx 2.0$, which is very substantial and should give a large residual $\kappa_{0}/T$, comparable to the case of line nodes \cite{Tannatar2010}.
Moreover, in the case of point nodes, the electronic specific heat in the superconducting state should follow a $T^3$ dependence \cite{Ran2019}, which contradicts the exponential behavior observed in 4Hb-TaS$_{2}$.
Therefore, point nodes in the superconducting gap are also unlikely.

The evolution of $\kappa_{0}/T$ with magnetic field can provide further information about the superconducting gap structure \cite{Shakeripour2009}.
Figure 4(b) illustrates the thermal conductivity of Sample A in magnetic fields up to 0.96 T ($\mu_{0} H \parallel c$).
For the data between 0 and 0.3 T,  the temperature dependence is still well described by the power law $\kappa/T = a + bT^\alpha$.
For magnetic fields of 0.5 T and above, the curves exhibit roughly linear behavior, making a linear fit of $\kappa/T = a+bT$ appropriate, i.e., fixing $\alpha=2$.
The dependence of $\alpha$ on the field indicates that the scattering of phonons by electrons in the vortex state becomes increasingly dominant, leading to a $T^2$ behavior of phonon thermal conductivity \cite{Etienne2003}.
Although the data are noisier, the normal-state $\kappa_{0}/T$ in 0.96 T roughly meets the expectation of Wiedemann-Franz law, $L_{0}/\rho_{0} (0.96 \rm T)=1.065$ mW K$^{-2}$  cm$^{-1}$.
The $\kappa_{0} (H)/T$ obtained in Fig. 4(b) is normalized to $\kappa_{\rm N0}/T$ and plotted as a function of $H/H_{\rm c2}$ in Fig. 4(c), with $\mu_{0} H_{\rm c2}=0.96$ T.
For comparison, similar data of the clean $s$-wave superconductor Nb \cite{Lowell1970}, the dirty $s$-wave superconducting alloy InBi \cite{PhysRevB.14.1916}, the multiband $s$-wave superconductor NbSe$_{2}$ \cite{Boaknin2003}, and an overdoped $d$-wave cuprate superconductor Tl-2201 \cite{Proust2002} are also plotted.

For the single-band $s$-wave superconductor like clean Nb \cite{Lowell1970} and dirty InBi \cite{PhysRevB.14.1916}, $\kappa_{0}/T$ is zero at $\mu_{0} H=0$ and increases exponentially at low fields, a phenomenon attributed to the tunnelling of normal-state quasiparticles between adjacent vortices.
In the nodal superconductor Tl-2201, a small field can induce rapid growth in the quasiparticle DOS due to the Volovik effect, with the low-field $\kappa_{0} (H)/T$ exhibiting a roughly $\sqrt{H}$ dependence  \cite{Proust2002}.
By contrast, in the typical multi-band $s$-wave superconductor NbSe$_{2}$, the magnitude of the gaps varies significantly across different sheets of the Fermi surface, therefore there is a more rapid increase in $\kappa_{0}/T$ at low fields and the overall field dependence of $\kappa_{0}/T$ relies on the ratio of different gap magnitudes \cite{Boaknin2003}.

In Fig. 4(c), the initial rise in $\kappa_{0} (H)/T$ versus $\mu_{0}H$ is slow and approximately exponential, indicating that field-induced quasiparticle heat conduction is an activated process in 4Hb-TaS$_{2}$.
This behavior corresponds to the suppression of a small full gap by the applied field, further confirming the absence of nodes in the gap.
In addition, $\kappa_{0} (H)/T$ shows an S-shaped dependence across the full field range, a hallmark of multiple $s$-wave superconducting gaps with different magnitudes, which was previously observed in NbSe$_{2}$, BaNi$_{2}$As$_{2}$ \cite{Kurita2009}, and TlNi$_{2}$Se$_{2}$ \cite{HXC2014}.
The multiple superconducting gaps are consistent with the multiple electronic bands found in 4Hb-TaS$_{2}$ \cite{Ribak2020,Almoalem2024}.

After ruling out the presence of nodal quasiparticles, the origin of the residual DOS in the superconducting state of 4Hb-TaS$_{2}$ remains puzzling.
A possible scenario is the gapless U(1) QSL state of the 1T layers in 4Hb-TaS$_{2}$.
Monolayer 1T-TaS$_{2}$ has been demonstrated to be a strongly correlated insulator and proposed as a candidate for the QSL with a spinon Fermi surface \cite{Patrick2018}. For monolayer 1T-TaSe$_{2}$ and 1T-NbSe$_{2}$, recent spectroscopy measurements also provide evidence for QSL state \cite{Ruan2021,NbSe22024}.
Neutral spinon excitations can carry heat and may give a residual linear term in thermal conductivity \cite{Nave2007}.
However, due to the significant difference in the work functions between the 1T and 1H layers of 4Hb-TaS$_{2}$, the existence of charge transfer from 1T to 1H layer has been widely acknowledged, which almost completely depletes the valence states in the 1T layers \cite{Beidenkopf2021,Almoalem2024,Nayak2023,Crippa2024}.
In this context, the intrinsic Mott physics is disrupted, ruling out the gapless QSL state in the 1T layers.

A more plausible scenario is that 4Hb-TaS$_{2}$ exhibits gapless superconductivity on certain Fermi surfaces.
Recent angle-resolved photoemission spectroscopy (ARPES) measurements have revealed multiple groups of Fermi surface sheets in 4Hb-TaS$_{2}$, including barrel-shaped pockets centered at the $\Gamma$ and K points, as well as dog-bone-shaped pockets around the M points, all derived from the 1H layer \cite{Yang2024,Almoalem2024}.
Additionally, a set of shallow electron pockets around the $\Gamma$ point has also been identified, associated with a narrow 1T-layer band slightly above the Fermi level \cite{Ribak2020,Wen2021}.
This indicates the presence of a small number of itinerant charge carriers in the 1T layer, consistent with the picture of a strongly doped Mott insulator \cite{Crippa2024}.
Therefore, 4Hb-TaS$_{2}$ can be regarded as a system composed of  alternating normal metal and superconducting layers.
Due to the intrinsic proximity effect and interband interactions, superconductivity should be induced in the 1T layer \cite{Kresin1992,Barzykin}.
However, STS measurements do not detect a superconducting gap in the 1T layer (at least in the top layer) \cite{Beidenkopf2021,Yan2022}.
Indeed, a relatively small number of magnetic impurities can completely suppress the induced superconducting gap in the intrinsically normal layer, leading to a gapless superconducting state, without significantly affecting the transition temperature \cite{Kresin1992}.
Note that the 1T-layer spectra show Kondo-like peaks on about 12\% of the CDW sites, indicating that, despite the breakdown of the Mott insulating state, a few local magnetic moments still persist in the 1T layer \cite{Nayak2023}.
Therefore, a reasonable speculation is that these residual local moments induce gapless superconductivity in the shallow pockets from the 1T layer, which could explain our thermal conductivity results.

On the other hand, STS measurements on the 1H layer also exhibit finite in-gap DOS \cite{Beidenkopf2021}.
A recent theory proposes that gapless superconductivity may extend into the 1H-layer pockets via interlayer hybridization \cite{Dentelski2021}.
Due to the strong band dependence of hybridization, gapless superconductivity likely emerges in those 1H-layer pockets strongly coupled with the 1T-layer local moments, while those nearly decoupled 1H-layer pockets remain fully gapped.
This interpretation is also compatible with our results.
More experiments, such as ultralow-temperature ARPES measurements, are highly desired to clarify the true superconducting ground state in 4Hb-TaS$_{2}$.

In summary, we have measured the ultralow-temperature thermal conductivity of 4Hb-TaS$_{2}$ single crystals.
A small residual linear term $\kappa_{0}/T$ in zero field and an S-shaped field dependence of $\kappa_{0}(H)/T$ are observed.
These behaviors suggest that 4Hb-TaS$_{2}$ is likely to simultaneously host both gapless Fermi pockets and fully gapped superconducting pockets, with the former contributing to the residual DOS and the latter exhibiting multiple superconducting gaps.
Our results impose constaints on the superconducting gap structure, and providing further clues for elucidating the pairing mechanisms in 4Hb-TaS$_{2}$.

This work was supported by the Natural Science Foundation of China (Grants No. 12174064 and No. 12034004), the Shanghai Municipal Science and Technology Major Project (Grant No. 2019SHZDZX01), and the Innovation Program for Quantum Science and Technology (Grant No. 2024ZD0300104). H.C.L. was supported by the Beijing Natural Science Foundation (Grant No. Z200005), the National Key R$\&$D Program of China (Grants No. 2022YFA1403800 and 2023YFA1406500), the Natural Science Foundation of China (Grant No. 12274459).

\end{document}